\documentclass[%
 reprint,
superscriptaddress,
 amsmath,amssymb,
 aps,
floatfix,
]{revtex4-1}
\usepackage[colorinlistoftodos]{todonotes}
\usepackage{graphicx}
\usepackage{dcolumn}
\usepackage{bm}
\usepackage[mathlines,pagewise]{lineno}
\usepackage{xcolor}

\usepackage{etoolbox}

\newcommand{\avg}[1]{ \left\langle #1 \right\rangle }

\begin{document}

\preprint{APS/123-QED}

\title{Entropy and Heat Capacity of the transverse momentum distribution for pp\\collisions at RHIC and LHC energies}

\author{D. Rosales Herrera, J. R. Alvarado Garc\'ia, A. Fern\'andez T\'ellez}
\affiliation{%
 Facultad de Ciencias Físico Matemáticas, Benemérita Universidad Autónoma de Puebla,\\ Apartado Postal 165, 72000 Puebla, Puebla, México
}%

\author{J. E. Ram\'irez}
\email{jhony.eredi.ramirez.cancino@cern.ch}
\affiliation{
 Centro de Agroecología, Instituto de Ciencias, Benemérita Universidad Autónoma de Puebla, Apartado Postal 165, 72000 Puebla, Puebla, M\'exico
}%
\author{C. Pajares}
\affiliation{Departamento de Física de Partículas and Instituto Galego de Física de Altas Enerxías, Universidad de Santiago de Compostela, E-15782 Santiago de Compostela, España
}%



\begin{abstract}
We investigate the transverse momentum distribution (TMD) statistics from three different theoretical approaches. In particular, we explore the framework used for string models, wherein the particle production is given by the Schwinger mechanism. The thermal distribution arises from the Gaussian fluctuations of the string tension. The hard part of the TMD can be reproduced by considering heavy tailed string tension fluctuations, for instance, the Tsallis $q$-Gaussian function, giving rise to a confluent hypergeometric function that fits the entire experimental TMD data. 
We also discuss the QCD-based Hagerdon function, another family of fitting functions frequently used to describe the spectrum.
We analyze the experimental data of minimum bias pp collisions reported by the BNL Relativistic Heavy Ion Collider (RHIC) and the CERN Large Hadron Collider (LHC) experiments (from $\sqrt{s}=0.2$ TeV to $\sqrt{s}=13$ TeV). We extracted the corresponding temperature by studying the behavior of the spectra at low transverse momentum values. 
For the three approaches, we compute all moments, highlighting the average, variance, and kurtosis. Finally, we compute the Shannon entropy and the heat capacity through the entropy derivative with respect to the temperature.
We found that the $q$-Gaussian string tension fluctuations lead to a monotonically increasing heat capacity as a function of the center of mass energy, which is also observed for the Hagedorn fitting function. This behavior is consistent with the experimental observation that the temperature slowly rises with increments of the collision energy.
\end{abstract}

\keywords{Transverse momentum spectra, Schwinger mechanism, Hagedorn, pp collisions}
\maketitle



\section{Introduction\protect\\}

The study of high-energy ion collisions has been a significant area of research in nuclear and particle physics, providing insights into the properties of strongly interacting matter under extreme conditions \cite{Bjorken:1982qr}. 
One relevant experimental measurement is the transverse momentum distribution (TMD), which is a \emph{histogram} built with the transverse momentum ($p_T$) of the produced charged particles per momentum space unit and contains information on the processes involved in all scales of events, leading to the final state of produced particles \cite{VOGT2007221}. 
The importance of the TMD renders the study of theoretical models and empirical fitting functions that adequately describe part or all the spectrum necessary.
Earlier efforts to achieve this assume that the TMD follows an exponential distribution, where the inverse of the exponential decay is frequently associated with the temperature of the collision system \cite{Becattini:1995if,becattini1997thermal}. 
This fitting function reasonably described the experimental data at the lower center of mass energies but deviated as experiments reached higher energies, revealing a non-exponential tail \cite{Bialas:2015pla,feal2021thermal}.
This approach is valid when most of the contribution to the spectra comes from soft scattering processes, leading to a soft thermal-like $p_T$ distribution \cite{Braun-Munzinger:1994ewq,Braun-Munzinger:2003htr,feal2021thermal}.

In the early `80s, Hagedorn introduced a QCD-based fitting function described by a power law of the transverse momentum shifted by a threshold that comes from the elastic scattering momentum scale \cite{Hagedorn:1965st,Hagedorn:1964zz,Hagedorn:1983wk}. Interestingly, this proposal reproduced both behaviors of the TMD, thermal and a power law tail at low and high $p_T$ values, respectively.
Later, the high-energy community presented a new fitting function based on the Tsallis $q$-Exponential function, which generalizes the thermal distribution by introducing a certain non-extensivity degree of the systems formed in high energy collisions \cite{PRLWilk, biro2020tsallis}.
However, these fitting functions are shown to be equivalent \cite{Saraswat:2017kpg}.

On the other hand, for string models, the production of charged particles is described by creating neutral color pairs through the breaking of the strings stretched between the partons. These subsequently decay, producing the observed hadrons \cite{andersson1998lund}.
In these cases, the transverse momentum distribution is governed by the Schwinger mechanism \cite{schwinger}. 

In the latter `90s, Bialas reconciled this approach with the thermal distribution by considering the string tension undergoes Gaussian fluctuations with zero average and variance proportional to the string tension \cite{BIALAS1999301}. Later, C. Pajares resumed this idea to incorporate a temperature-like parameter on the Color String Percolation Model, and thus, he provided a way to compute the string density from experimental data \cite{DiasdeDeus:2006xk,Braun:2015eoa,bautista2019string}.
Recently, in Refs.~\cite{Pajares:2022uts,Garcia:2022eqg}, the authors retake the original Bialas idea to extend the string tension fluctuations to a heavy tailed distribution. In particular, if the tensions fluctuate according to a $q$-Gaussian distribution, then the TMD becomes a confluent hypergeometric function that correctly fits the spectrum, thermal and power law behaviors at low and high $p_T$ values, respectively \cite{Pajares:2022uts,Garcia:2022eqg}.
 
In this manuscript, we analyze the TMD data of minimum bias pp collisions reported by the experiments at the Relativistic Heavy-Ion Collider (RHIC), BNL and the Large Hadron Collider (LHC), CERN under the three schemes discussed above, namely, the thermal distribution, the Hagedorn, and the confluent hypergeometric fitting functions. 
In this way, we discussed the thermal temperature estimated in each scenario as a function of the center of mass energy.
Since each fitting function has a different degree of accuracy in reproducing the spectrum, we compute some statistics to compare them, such as the average of transverse momentum, variance, and kurtosis.
Additionally, we compute the Shannon entropy for each fitting function and estimate the heat capacity.
The latter determination helps estimate the energy increment necessary to heat the collision systems.

The plan of the paper is as follows. In Sec.~\ref{sec:TMD}, we comment on different approaches to describe the TMD and their main features. In Sec.~\ref{sec:TMDfit}, we show the fits to the TMD of minimum bias pp collisions and give a description of fit parameters as a function of the center of mass energy. In Sec.~\ref{sec:moments}, we compute the moments of the TMD in the approaches discussed in Sec.~\ref{sec:TMD}.
Section \ref{sec:Shannon} contains our computations of the Shannon entropy and the heat capacity of the analyzed TMD data.
Finally, in Sec.~\ref{sec:conclusions}, we write our final comments, conclusions, and perspectives.

\section{Theoretical description of the TMD}\label{sec:TMD}

In this section, we discuss the particularities of the transverse momentum distribution, which can be obtained from different approaches. In particular, we are interested in discussing the cases of the TMD for color string systems and the QCD-based fitting function proposed by Hagedorn. 
In what follows, the TMD is denoted as $dN/dp_T^2$, meaning the invariant yield of produced particles.

\subsection{TMD from the Schwinger mechanism}\label{ssec:SM}

The Schwinger Mechanism explains the generation of particle-antiparticle pairs from the quantum vacuum under the influence of an intense gauge field, which supplies the necessary energy to convert the field's energy into particle neutral pairs. This phenomenon occurs upon the gauge field surpassing a certain critical intensity, enabling the field's energy to materialize as mass \cite{schwinger}.

In terms of the resultant particle dynamics, the Schwinger Mechanism describes a Gaussian behavior in the TMD of the produced particles. This behavior arises due to the exponential damping linked to the energy barrier imposed by the vacuum \cite{bialas1986conversion}. As the transverse momentum magnifies, the likelihood of particle-antiparticle pair production diminishes exponentially. 
Initially, the Schwinger Mechanism was conceived to explain the emergence of electron-positron pairs within a potent electromagnetic field \cite{schwinger}; and then, it was broadened to the creation of quark-antiquark and quark-quark--antiquark-antiquark pairs within the framework of QCD. These pairs promptly amalgamate into color-neutral hadrons, yielding the transverse momentum distribution of the observed particles.

Taking these ideas into consideration, the probability of observing a produced hadron with momentum $p_T$ is proportional to \cite{schwinger,wong1994introduction}:
\begin{equation}
    \frac{dN}{dp_T^2}\sim e^{ -\pi p_T^2/x^2},
    \label{eq:SM}
\end{equation}
where $x^2$ is the string tension associated with the energy supplied to the vacuum in the context of QCD color string models. 

\subsection{Thermal TMD}\label{ssec:thermal}

As we commented in Sec.~\ref{ssec:SM}, the Schwinger mechanism has been adequately adapted to describe the production of charged particles in high energy physics and relates the energy supplied to the vacuum for the pair creation with the string tension in QCD \cite{Andersson:1983ia}.
First, it was proposed that the tension of the color string was taken as a constant.
Later, Bialas introduced the string tension fluctuations based on a stochastic QCD-vacuum approach \cite{BIALAS1999301}. 
In this way, if the tension of the strings is considered as a random variable described by a probability density function $P(x)$, then the appropriate computation of the spectrum should consider these fluctuations, which can be done by performing the following convolution
\begin{equation}
    \frac{dN}{dp_T^2}\propto \int_0^\infty e^{-\pi p_T^2/x^2} P(x)dx.
\label{eq:convolution}
\end{equation}
Assuming that the string tension fluctuations are described by a Gaussian distribution, the Schwinger mechanism becomes 
\begin{equation}
    \frac{dN}{dp_T^2} \propto e^{-p_T/T_\text{th}},
    \label{eq:thermal}
\end{equation}
where $T_\text{th}={\avg{p_T}}/2$ \cite{BIALAS1999301}. 
Equation~\eqref{eq:thermal} can be interpreted as a \emph{thermal distribution} because it is similar to the Boltzmann distribution.
$T_\text{th}$ can be understood as the temperature linked to the TMD, computed over the ensemble of collision events occurring under identical conditions \cite{kubo2012statistical}.

\begin{figure*}[ht]
    \centering     
    \includegraphics[scale=1]{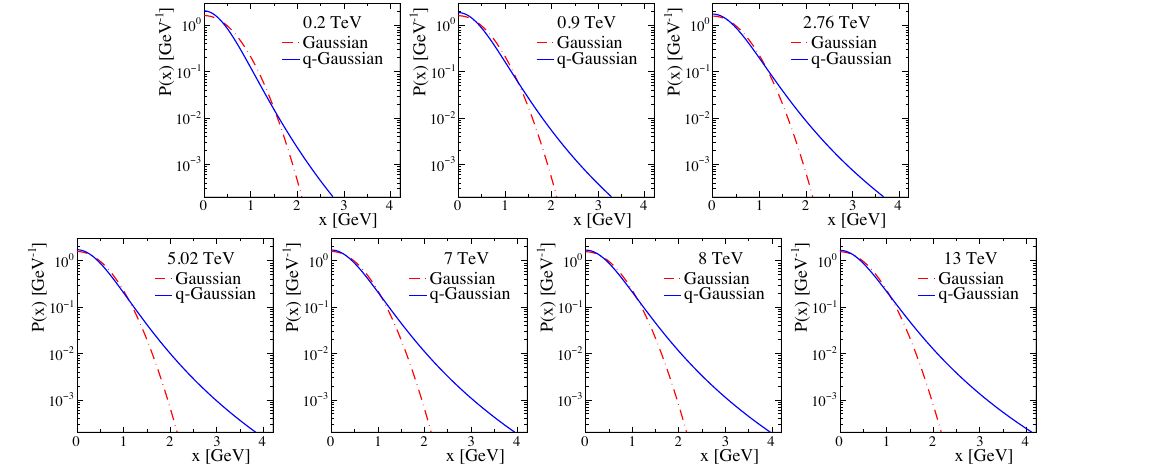}      
    \caption{Gaussian and $q$-Gaussian string tension fluctuations for pp collisions at different center of mass energies. The tail of the $q$-Gaussian distribution becomes less tilted as the center of mass energy increases, which means that the probability of observing a string with a higher tension rises.}
    \label{fig:fluctuations}
\end{figure*}

\subsection{Soft and hard scales of the TMD from string tension fluctuations}\label{ssec:U}

In Sec.~\ref{ssec:thermal}, we discussed the origin of the TMD thermal distribution from the fluctuations of the string tension. This approach reproduces the thermal behavior of the TMD but fails to describe the power-like law of the TMD tail at high $p_T$ values.

Recently, it has been shown that the spectrum can be adequately described by replacing the Gaussian fluctuations of the string tension with a $q$-Gaussian distribution \cite{Pajares:2022uts,Garcia:2022eqg}, which reads
\begin{equation}
P(x)=\mathcal{N}\left( 1+\frac{(q-1)x^2}{2\sigma^2}  \right)^\frac{1}{1-q},
\label{eq:qGfluct}
\end{equation}
with zero mean value, scale parameter $\sigma$, and $\mathcal{N}$ being the normalization constant.
To allow variations in the string tension across a range from zero to infinity, it is necessary that the parameter $q$ lies between 1 and 3 \cite{budini2015extended,budini2,Garcia:2022eqg}.
In general, this $q$-Gaussian distribution is heavy tailed, which means that in the collision system, the probability of observing a string with higher tensions is greater than in the Gaussian case, as we depict in Fig.~\ref{fig:fluctuations} for pp collisions at different center of mass energies (further analyzed below).

Following the same procedure as in Sec.~\ref{ssec:thermal}, and by introducing the variable $t = 2\sigma^2 / [(q-1)x^2]$,
the convolution~\eqref{eq:convolution} now reads
\begin{equation}
    \frac{dN}{dp_T^2} \propto \int_0^\infty e^{- \frac{\pi p_T^2 (q-1)}{2\sigma^2} t} 
    t^{\frac{1}{q-1} - \frac{3}{2}} 
    (1+t)^{\frac{1}{q-1}} dt.
\label{eq:convU}
\end{equation}
By comparing Eq.~\eqref{eq:convU} with the confluent hypergeometric function defined as
\begin{equation}
    U(a, b, z)=  \frac{1}{\Gamma(a)}\int_0^\infty e^{-zt} t^{a-1}(1+t)^{b-a-1} dt,
    \label{eq:U}
\end{equation}
we identify 
\begin{equation}
a =  \frac{1}{q-1}-\frac{1}{2}, \quad
b  =  1/2, \quad \text{and} \quad
z  =  \pi p_T^2 \frac{q-1}{2\sigma^2}.
\label{eq:abz}
\end{equation} Therefore, the spectrum~\eqref{eq:convU} becomes
\begin{equation}
\frac{dN}{dp_T^2}\sim  \Gamma \left( \frac{1}{q-1}  \right) U\left( \frac{1}{q-1}-\frac{1}{2}, \frac{1}{2}, \pi p_T^2 \frac{q-1}{2\sigma^2}  \right).
\label{eq:U2}
\end{equation}

In particular, Eq.~\eqref{eq:U2} reproduces the exponential decay at low $p_T$ region, reaching the asymptotic behavior 
\begin{equation}
    \frac{dN}{dp_T^2} \sim e^{ - p_T/T_U  },
    \label{eq:U2thermal}
\end{equation}
where the thermal temperature $T_U$ is defined as
\begin{equation}
    T_U=\sigma \frac{\Gamma\left( \frac{1}{q-1}-\frac{1}{2} \right)}{\sqrt{2\pi (q-1)}\Gamma\left( \frac{1}{q-1} \right)}. 
    \label{eq:Tthermal}
\end{equation}

On the other hand, for high $p_T$ values, the TMD \eqref{eq:U2} behaves as a power law in $p_T^2$
\begin{equation}
    \frac{dN}{dp_T^2} \sim 
    \frac{\Gamma\left( \frac{1}{q-1} \right)}{\sqrt{\pi}} \left(  \frac{\pi p_T^2 (q-1)}{2\sigma^2} \right)^{\frac{1}{2}-\frac{1}{q-1}}, 
    \label{eq:U2hard}
\end{equation}
from where we define the hard scale
\begin{equation}
    T_{H,U}= \sigma \sqrt{\frac{2}{\pi(q-1)}} \left( \frac{\sqrt{\pi}}{\Gamma\left( \frac{1}{q-1} \right)}  \right)^{\frac{q-1}{q-3}}.
    \label{eq:Thard}
\end{equation}
Note that the soft~\eqref{eq:Tthermal} and hard~\eqref{eq:Thard} scales only depend on $q$-Gaussian parameters.

It is worth mentioning that the $U$ fitting function reproduces the well-known soft and hard behaviors of the TMD \cite{U}.

\subsection{Hagedorn-like fitting functions}

A different description of the TMD comes from a formula inspired by QCD. The Hagedorn distribution is concerned with describing the hard part of the spectrum at high $p_T$ values. This fitting function is given by
\begin{equation}
      \frac{dN}{dp_T^2} \propto \left(\frac{p_0}{p_0 + p_T}\right)^m. 
    \label{eq:Hag1}
\end{equation}
It is straightforward to show that the latter is a Tsallis $q$-Exponential function by doing the following parametrization: $m= 1/(q_e-1)$ and $p_0 = \lambda/(q_e-1)$. Therefore, the Hagedorn spectrum becomes
\begin{equation}
    \frac{dN}{dp_T^2} \propto \left( 1 + \frac{p_T}{p_0} \right)^{-m} = \left( 1 + (q_e-1) \frac{p_T}{\lambda} \right)^{\frac{1}{1-q_e}},
\label{eq:Hag2}\end{equation}
where we have added the subscript $e$ to $q_e$ in order to avoid misleading on the $q$ parameter of the string tension fluctuations~\eqref{eq:qGfluct}.

Other authors consider a similar fitting function, which corresponds to \eqref{eq:Hag2} to the $q_e$ for thermodynamic consistency where the variable is the transverse mass $m_T^2=p_T^2+m_0^2$, and $m_0$ is the mass of the produced particle \cite{Cleymans:2011ij,Parvan:2019aii,Sahu:2021cdl,Tao:2022tcw}. The Tsallis distribution has been used to fit the TMD by the experiments such as the STAR collaboration \cite{STAR:2006nmo} at RHIC, and by the ALICE \cite{ALICE:2011gmo} and CMS \cite{CMS:2011jlm} collaborations at LHC. It is convenient to replace $m_T$ with $p_T$. In this case, the TMD is given by
\begin{equation} 
     \frac{dN}{dp_T^2} \propto \left(1 + (q_e'-1)\frac{p_T}{\tau} \right)^{\frac{q_e'}{1-q_e'}},
\label{eq:Hag3}\end{equation}
which also can be expressed as a Tsallis $q$-Exponential if we replace $q_e$ and $\lambda$ with $2 - 1/q_e'$ and $\tau/q_e'$, respectively.
Notice that Eqs.~\eqref{eq:Hag1}, \eqref{eq:Hag2}, and \eqref{eq:Hag3} are equivalent, and they must describe the same behaviors of the TMD at low and high $p_T$ values. For instance, in the limit of low $p_T$
\begin{equation}
    \frac{dN}{dp_T^2} \propto 1-\frac{p_T}{T_\text{Hag}} + \mathcal{O}(p_T^2) \sim  e^{-p_T/T_\text{Hag}},
\end{equation}
where $T_\text{Hag}$ is the slope of the spectrum at low $p_T$, which is given by $p_0/m$, $\lambda$, and $\tau/q_e'$ for~\eqref{eq:Hag1},~\eqref{eq:Hag2}, and~\eqref{eq:Hag3}, respectively. 
Nevertheless, it is worth mentioning that, for the cases of the $q$-Exponential~\eqref{eq:Hag3},  $\tau$ is usually used as a temperature parameter \cite{Cleymans:2011ij,Sahu:2021cdl,Tao:2022tcw}. 
However, this parameter does not consider the complete slope in the argument of the thermal distribution, leading to a subestimation of thermal temperature when $q_e' > 1$, as observed in the cases discussed in this manuscript. 

On the other hand, at high values of $p_T$, it is found that 
\begin{equation}
      \frac{dN}{dp_T^2} \propto \left( \frac{p_T}{T_{H,\text{Hag}}}   \right)^\alpha,
\label{eq:HagH}
\end{equation}
where $T_{H,\text{Hag}}$ is identified in~\eqref{eq:HagH} as $p_0$ \cite{Hagedorn:1983wk}, $\lambda/(q_e - 1)$, and $\tau/(q_e' - 1)$ for~\eqref{eq:Hag1},~\eqref{eq:Hag2}, and~\eqref{eq:Hag3}, respectively. Usually, $T_{H,\text{Hag}}$ is named as the hard scale of the TMD \cite{Bylinkin,Bylinkin:2014fta,Baker:2017wtt, Feal:2018p, Bellwied_2018}. Moreover, the $\alpha$ parameter corresponds to the exponent of the TMD fitting function of each case. 

\section{Experimental TMD data analysis}\label{sec:TMDfit}

\begin{figure*}[ht]   
    \includegraphics[scale=1]{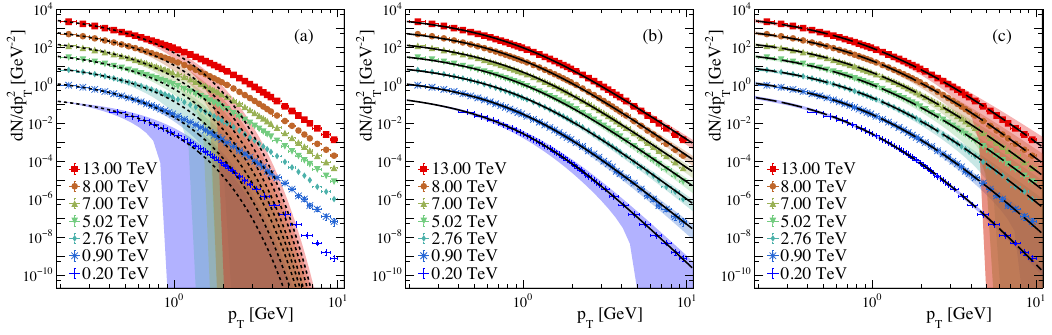} 
\caption{Fits (lines) to the experimental data (figures) for minimum bias pp collisions at different center of mass energies by using (a) the thermal distribution, (b) the hypergeometric confluent $U$ function, and (c) the Hagedorn function. Shaded regions correspond to the 95\% uncertainty propagation.}
\label{fig:fits}
\end{figure*}
    \vspace{-5pt}

\begin{table}[ht]
    \caption{$p_T$ range for the thermal fit and their corresponding temperatures.}
            \label{tab:therfitpar}
    \centering
    \begin{ruledtabular}
    \begin{tabular}{ c c c c } 
       $\sqrt{s}$ [TeV] & $p_{T,\text{min}}$ [GeV] & $p_{T,\text{max}}$ [GeV] & $T_\text{th}$ [GeV] \\ \hline
0.20	&	0.40	&	0.90	&	0.197(31)	\\
0.90	&	0.15	&	0.70	&	0.199(10)	\\
2.76	&	0.15	&	0.50	&	0.202(20)	\\
5.02	&	0.15	&	0.60	&	0.203(13)	\\
7.00	&	0.15	&	0.55	&	0.202(16)	\\
8.00	&	0.15	&	0.60	&	0.204(14)	\\
13.00	&	0.15	&	0.60	&	0.205(13)	\\ 
    \end{tabular}
    \end{ruledtabular}
\end{table}

\begin{figure}[ht]
\includegraphics[scale=0.25]{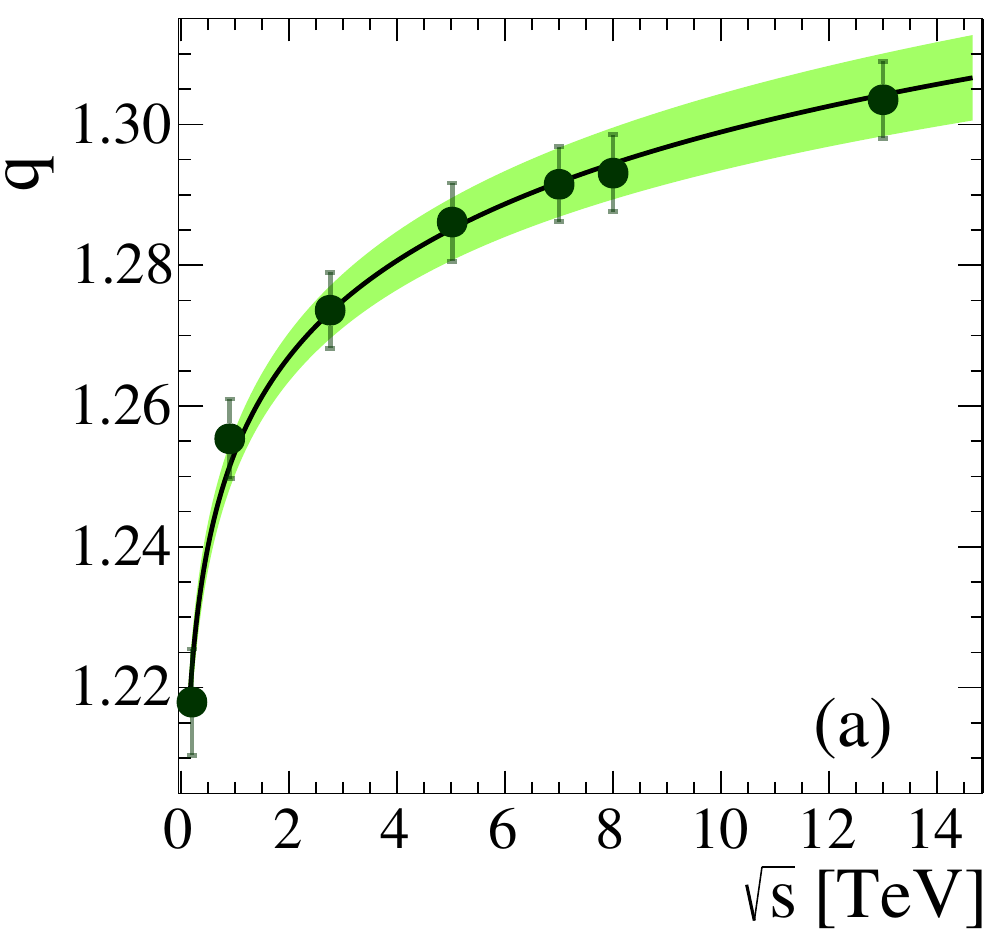}
\includegraphics[scale=0.25]{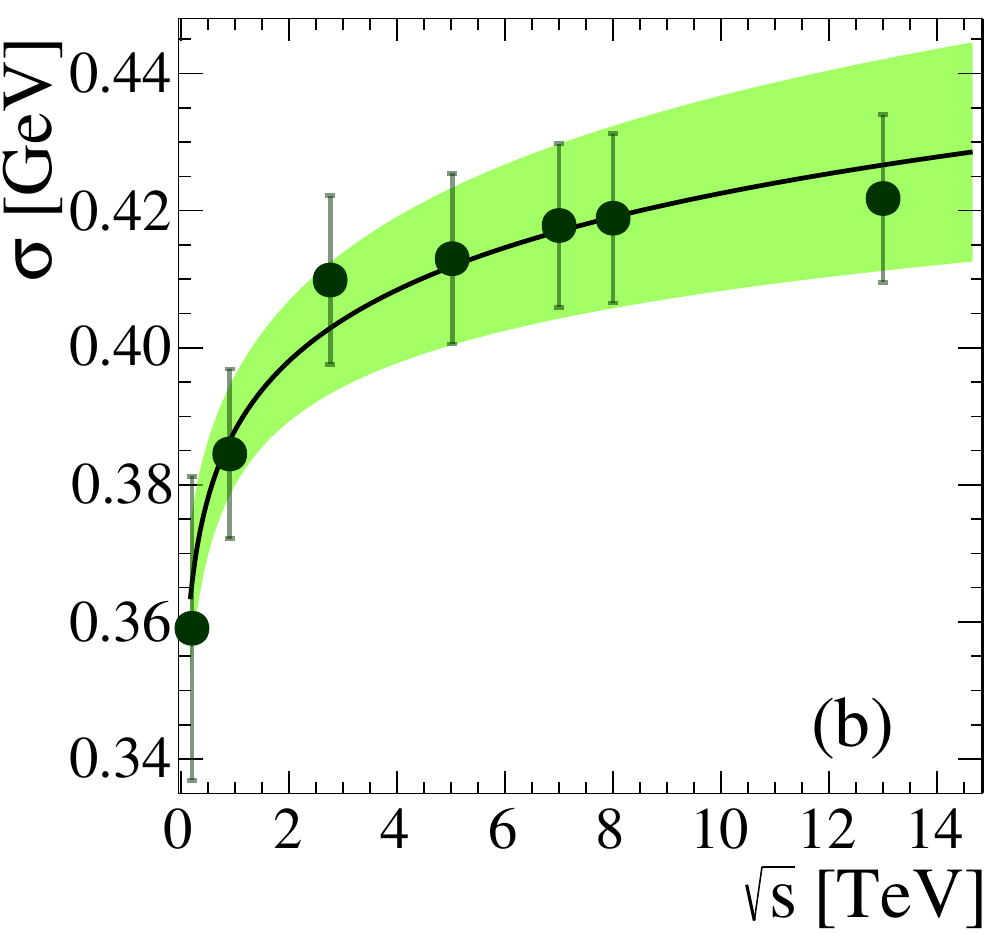}
\caption{Values of the parameters (a) $q$ and (b) $\sigma$ obtained from the confluent hypergeometric $U$ fits as a function of center of mass energy for minimum bias pp collisions. The scaling law dependences \eqref{eq:allfit} are shown in solid lines. Shaded regions correspond to the uncertainty propagation. }
\label{fig:sfit}
\end{figure}

We analyze the experimental transverse momentum spectra of charged particles of minimum bias pp collisions at different center of mass energies. By using Eqs.~\eqref{eq:thermal},~\eqref{eq:U2}, and~\eqref{eq:Hag1}, we fit over the experimental data reported on Refs.~\cite{STAR:2003fka,ALICE:2010syw,ALICE:2022xip} using the ROOT 6 software. 
The fits were performed by using different $p_T$ ranges. For instance, we adjust the $p_T$ range for the thermal fits, finding the minimization of $\chi^2$ (see Table~\ref{tab:therfitpar}). However, for the Hagedorn and $U$ functions, the fit was done for the entire $p_T$ range reported by the experiments \cite{STAR:2003fka,ALICE:2010syw,ALICE:2022xip}.
In all cases, the value of the quotient $\chi^2$/NDF does not exceed 1 for the fits performed to the TMD data. Nevertheless, $\chi^2$/NDF$\gg$1 in the case of the thermal fits extrapolated to the entire range of $p_T$. This means that the three functions can provide a good description of the experimental data in the appropriate $p_T$ range.
As seen in Fig.~\ref{fig:fits}, each fitting function describes part of the spectrum: the thermal fit reproduces only the low $p_T$ region. Meanwhile, the Hagedorn was proposed to describe the high $p_T$ region of the spectrum (0.3 GeV $< p_T <$ 10 GeV \cite{Hagedorn:1983wk}) but is capable of reproducing the complete range of experimental data. Finally, the confluent hypergeometric confluent function successfully describes the behavior of the whole TMD for the data sets.

It is found that the $q$-Gaussian parameters rise as the center of mass energy increases. We propose the following scaling laws to describe these behaviors
\begin{subequations}  
\label{eq:allfit}
\begin{eqnarray}
     q(\sqrt{s}\text{ })
    &=&
    a_q   \left( \sqrt{\frac{s}{s_0}}\text{ } \right)^{c_q} , 
\label{eq:qfit}\\    
 \sigma(\sqrt{s}\text{ })  
&=&
a_{\sigma} \left( \sqrt{\frac{s}{s_0}}\text{ }  \right)^{c_{\sigma}} ,
\label{eq:sigmafit}  
\end{eqnarray}
\end{subequations}
with $\sqrt{s_0}= 1$ GeV, $a_q=1.253(3)$, $c_q=0.0154(14)$, $a_\sigma=0.388(8)$ GeV, and $c_\sigma=0.037(11)$. In Fig.~\ref{fig:sfit} we can see the $q$, and $\sigma$ dependence with $\sqrt{s}$ described by Eqs.~\eqref{eq:allfit}. 

\begin{figure}[ht]
\includegraphics[scale=0.25]{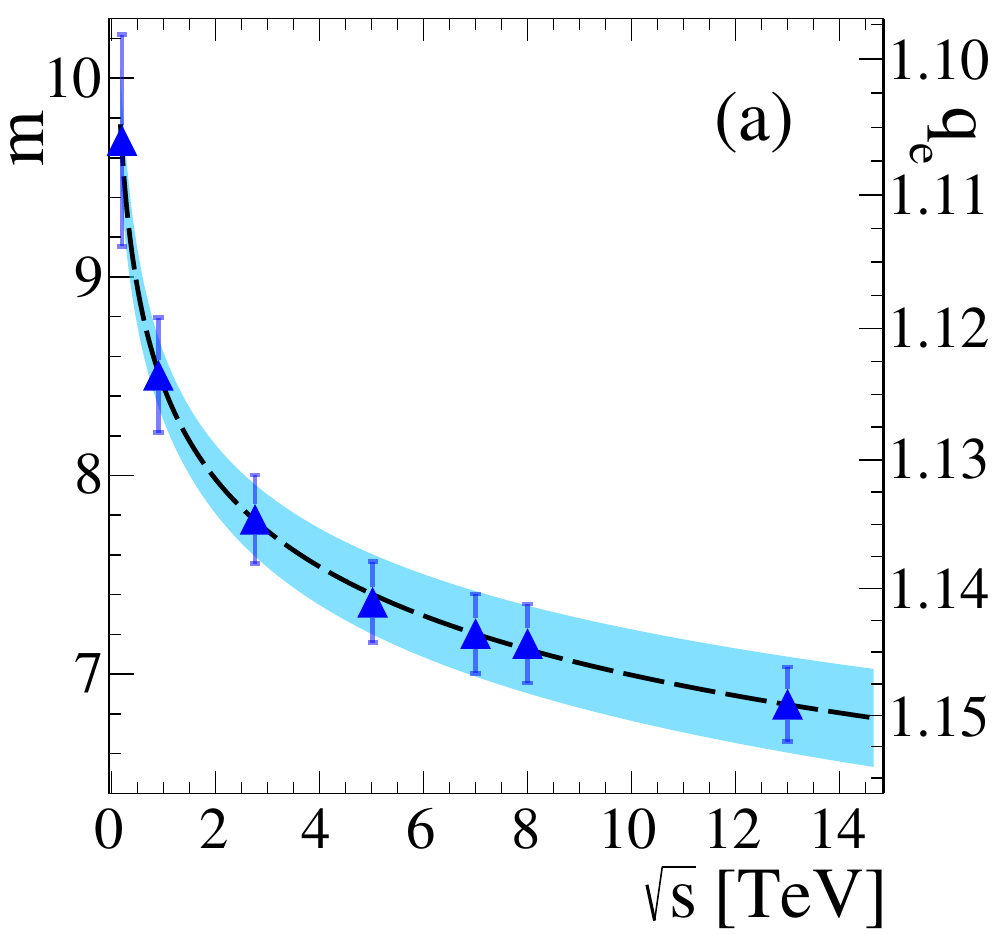}
\includegraphics[scale=0.25]{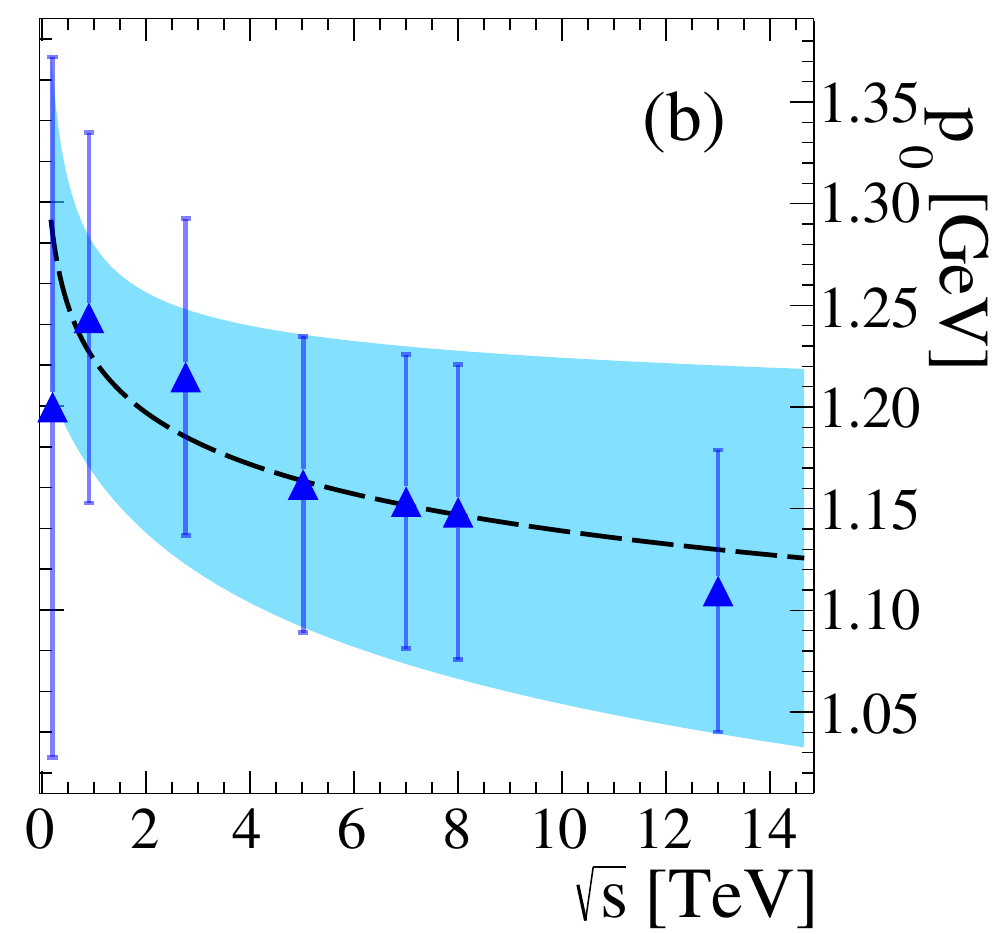}
\caption{Parameters (a) $m$ and (b) $p_0$ obtained from Hagedorn fits \eqref{eq:Hag1} to minimum bias pp collisions data as a function of the center of mass energy. Dashed lines are the parameters described by \eqref{eq:Hagpar}. The shaded regions correspond to the propagation of uncertainty.}
\label{fig:hfit}
\end{figure}

For the Hagedorn function, the fitting parameters $m$ and $p_0$ are described by: 
\begin{subequations}
\label{eq:Hagpar}
    \begin{eqnarray}
     m(\sqrt{s}\text{ }) &=& a_m \left( \sqrt{\frac{s}{s_0}}\text{ }  \right)^{c_m} ,
\label{eq:mfit}   \\
     p_0(\sqrt{s}\text{ }) &=& a_{p_0} \left( \sqrt{\frac{s}{s_0}}\text{ }  \right)^{c_{p_0}} ,
\label{eq:p0fit}    
\end{eqnarray}
\end{subequations}
with $a_m  =8.45(17)$,  $c_m = -0.082(11)$,  $a_{p_0}=1.22(5)$ GeV, and $c_{p_0}= -0.03(3)$.
Fig.~\ref{fig:hfit} shows the Hagedorn parameters dependence with $\sqrt{s}$. The parameter $q_e =(1+m)/m$ is also shown in Fig.~\ref{fig:hfit}(a) in the y-axis on the right side.

\begin{figure}[ht]
\includegraphics[scale=0.5]{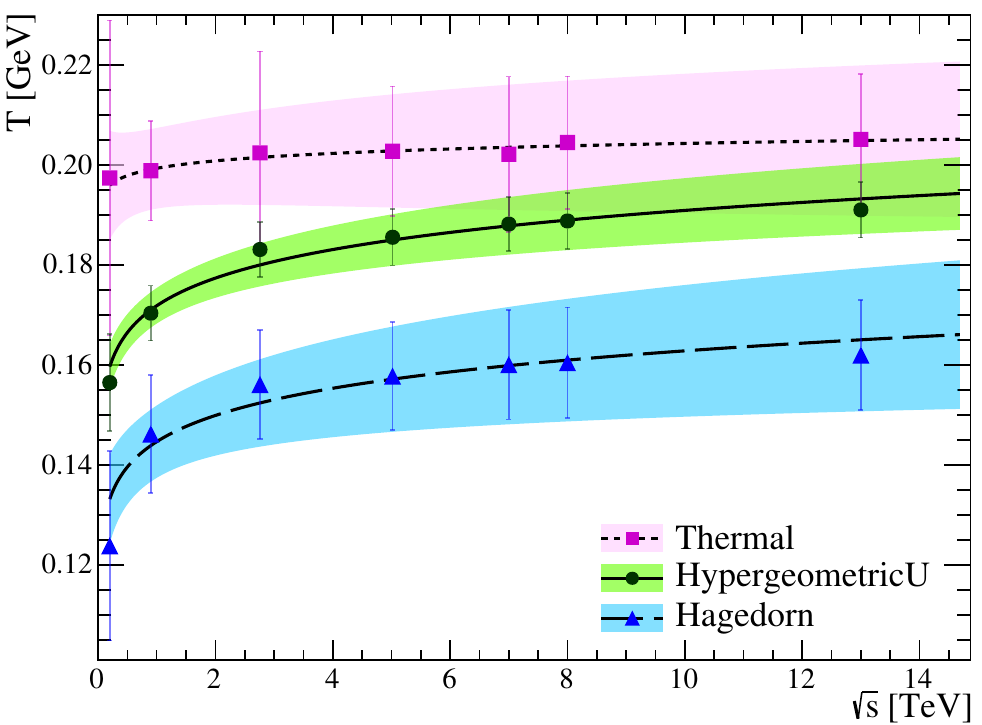}
\caption{Temperature extracted from (squares) Thermal, (circles) confluent hypergeometric $U$, and (triangles) Hagedorn fitting functions. Lines are the trend of the (dotted) $T_\text{th}$, (solid) $T_U$, and (dashed) $T_\text{Hag}$ described by Eq.~\eqref{eq:Tfit}. Shaded regions correspond to the uncertainty propagation.}
\label{fig:Temp}
\end{figure}

We recall that the three different approaches to analyzing the TMD provide their estimation of the thermal temperature. We found that $T_\text{th}$, $T_U$, and $T_\text{Hag}$ scale with the center of mass energy as \cite{Garcia:2022eqg}
\begin{equation}  
T(\sqrt{s}\text{ }) 
    = 
    a_{T} \left( \sqrt{\frac{s}{s_0}}\text{ }  \right)^{c_{T}} .
\label{eq:Tfit}
\end{equation}
The obtained parameters of Eq.~\eqref{eq:Tfit} for the three approaches are shown in Table~\ref{tab:temparam}, and they are plotted in Fig.~\ref{fig:Temp}.

\begin{table}[ht]
    \caption{Fit parameter values of the temperature behavior as a function of the center of mass energy \eqref{eq:Tfit} for each model.}    
        \label{tab:temparam}
    \begin{ruledtabular}
    \centering    
    \begin{tabular}{ccc} 
      Model   &  $a_T$ [GeV] & $c_T$ \\ \hline
      Thermal   &  0.199(8) & 0.011(24) \\ 
      HypergeometricU   &  0.172(4) & 0.046(11) \\
      Hagedorn   &  0.145(7) & 0.051(28) \\      
             \end{tabular}
    \end{ruledtabular}            
\end{table}

We recall that the temperature parameter is given by the TMD behavior at low $p_T$. 
The experimental TMD data exhibit a thermal behavior at low $p_T$ for identified species of produced particles, including the Higgs boson \cite{Pajares:2022uts}. The higher the masses, the higher temperature values are expected. In the collisions are also produced heavy resonances which decay in lower mass particles, enhancing the low $p_T$ spectrum, but they can not be considered as formerly produced by the fragmentation of color string clusters.
Notice that the parameters $q$ and $m$ control the tail of the TMD, exhibiting a monotonic behavior with the center of mass energy, as shown in Fig.~\ref{fig:sfit}(a) and \ref{fig:hfit}(a). Similar behaviors are expected as a function of the multiplicity. Additionally, in the heavy-tailed string tension fluctuations approach, the high $p_T$ particle production can be considered as rare events, including jets. This information is implicitly incorporated in the tail of the $q$-Gaussian fluctuations. Nevertheless, this approach is not able to distinguish the longitudinal and transverse jet structure.

\section{Moments of the TMD} \label{sec:moments}
We compute the $n$-th moment of the transverse momentum spectra in the standard form
\begin{equation}
\langle \mathcal{P}_T^n \rangle= \frac{ \int_0^{\infty}  p_T^n  \text{TMD} dp_T}{\int_0^{\infty} \text{TMD} dp_T},
\label{eq:ptn}
\end{equation}
for all the fitting functions discussed in Sec.~\ref{sec:TMD}.
Here, we have introduced the notation $\avg{\mathcal{P}_T^n}$ to avoid misinterpretation with the computation of the moments reported by the HEP community. In those cases, the TMD must be integrated by considering the differential contributions of the longitudinal momentum component \cite{Hagedorn:1983wk,Bylinkin:2014fta}. Then 
\begin{equation}
    \avg{p_T^n}=\frac{\int p_T^n \text{TMD} 2\pi p_T dp_T}{\int \text{TMD} 2\pi p_T dp_T} =\frac{\avg{\mathcal{P}_T^{n+1}}}{\avg{\mathcal{P}_T}}. \label{eq:momentslab}
\end{equation}
The latter definition is also equivalent to considering $p_T^2$ as the random variable.

The calculation of \eqref{eq:ptn} is immediate for the thermal distribution, which gives $\langle \mathcal{P}_T^n \rangle_\text{exp}=n!T^n$.

Let us explain the computation of $\langle \mathcal{P}_T^n \rangle$ for the Hagedorn and confluent hypergeometric function $U$ in detail.
In both cases, we define the auxiliary function
\begin{equation}
    I_n=\int_0^{\infty}  p_T^n  \text{TMD} dp_T.
\label{eq:In}
\end{equation}
In this way, $\langle \mathcal{P}_T^n \rangle=I_n/I_0$.
For the Hagedorn function \eqref{eq:Hag1}, we found
\begin{eqnarray}
I_n &=& \int_0^{\infty}  p_T^n \left( 1+\frac{p_T}{p_0}  \right)^{-m} dp_T \nonumber\\
 &=& p_0^{n+1} \int_0^{1}  y^{m-n-2} \left( 1- y  \right)^{n} dy \nonumber\\
 &=& p_0^{n+1} B(m-n-1, n+1), \nonumber 
\end{eqnarray}
where $B$ is the Beta function, which is well defined for $m>n+1$.
 Therefore
\begin{equation}
    \langle \mathcal{P}_T^n \rangle_\text{Hag}= (m-1) p_0^n B(  m-n-1, n+1 ). 
\label{eq:nptHag}
\end{equation}

Similarly, for the $U$ function, we need to compute the integral
\begin{equation}
    I_n =
    \int_0^{\infty} p_T^n U(a, b, z) dp_T, \nonumber
\label{eq:InU}
\end{equation}
where $a$, $b$, and $z$ corresponds to Eq.~\eqref{eq:abz}.
By following the definition of the confluent hypergeometric function~\eqref{eq:U},
we rewrite $I_n$ as
\begin{equation}
I_n = 
\int_0^\infty \int_0^\infty p_T^n e^{ -\pi p_T^2 \frac{q-1}{2\sigma^2}t } t^{a-1}(1+t)^{b-a-1} dt dp_T.
\label{eq:Int}
\end{equation}
To simplify notation, we prescinded writing the factor $1/\Gamma(a)$, which appears in the denominator and numerator of Eq.~\eqref{eq:ptn}. Note that the integral over $p_T$ in Eq.~\eqref{eq:Int} is a Gaussian integral
\begin{equation}
    \begin{split}
     & \int_0^\infty p_T^n e^{ -\pi p_T^2 \frac{q-1}{2\sigma^2}t} dp_T 
\\ =& \frac{1}{2} \Gamma \left( \frac{n+1}{2} \right) \left( \pi \frac{q-1}{2\sigma^2} \right)^{-(n+1)/2} t^{-(n+1)/2}.       
    \end{split}
\end{equation}

By plugging the later on Eq.~\eqref{eq:Int} and performing the change of variable $y= (t+1)^{-1}$, the remaining integral becomes
\begin{equation}
    \begin{split}
 & \int_0^1 (1-y)^{a-1-(n+1)/2}y^{-(b-(n+1)/2)} dy  \\
=& B\left( 1- \left( b -\frac{n+1}{2}  \right), a-\frac{n+1}{2}\right).
    \end{split}
\label{eq:intt}    
\end{equation}

Finally, the $I_n$ integrals are given by
\begin{equation}
\begin{split}
  I_n= &\frac{1}{2} \Gamma \left( \frac{n+1}{2} \right) \left( \pi \frac{q-1}{2\sigma^2} \right)^{-(n+1)/2} 
  \\ &\times B \left( \frac{n+2}{2}, \frac{1}{q-1}- \frac{n+2}{2} \right), 
\end{split}
\label{eq:Insol}
\end{equation}
which are well defined if $q<({4+n})/({2+n}).$
So, the moments of the distribution are expressed as
\begin{equation}
\begin{split}
\langle \mathcal{P}_T^n \rangle_U= &\frac{1}{\sqrt{\pi}} \Gamma \left( \frac{n+1}{2} \right) \left( \frac{2-q}{q-1} \right) \left( \frac{2\sigma^2}{\pi (q-1)}  \right)^{n/2} 
  \\ & \times B \left( \frac{n+2}{2}, \frac{1}{q-1}- \frac{n+2}{2} \right).
\label{eq:nptU}
\end{split}
\end{equation}

It is worth mentioning that some experiments report the TMD without the functional normalization by dividing by $p_T$. In those cases, the function describing the transverse momentum spectra is $p_T (dN/dp_T^2)$, and the moments of the distribution are calculated as discussed above.
Notice that the moments \eqref{eq:momentslab} can also be expressed in terms of the $I_n$ integrals as $I_{n+1}/I_1$. Moreover, the ratio $\avg{\mathcal{P}_T^n}/\avg{ p_T^n}$ is given by
\begin{subequations}
\label{eq:ptnratio}    
\begin{eqnarray}
     \frac{\avg{\mathcal{P}_T^n}_\text{th}}{\avg{p_T^n}_\text{th}} &=& \frac{1}{n+1},
    \label{eq:ptnth}
   \\ 
        \frac{\avg{\mathcal{P}_T^n}_\text{Hag}}{\avg{p_T^n}_\text{Hag}} &=& \frac{m-n-2}{(n+1)(m-2)},
        \label{eq:ptnHag}
   \\ 
  \frac{\avg{\mathcal{P}_T^n}_U}{\avg{p_T^n}_U} &=& 
\frac{   B \left( \frac{1}{q-1} -\frac{3}{2} , \frac{1}{q-1}-\frac{n  + 2}{2} \right)}{(n+1)  B \left( \frac{1}{q-1} -1 , \frac{1}{q-1}-\frac{n + 3}{2}\right)},  \qquad 
    \label{eq:ptnU}
    \end{eqnarray}    
\end{subequations}
for the thermal distribution, Hagedorn, and the confluent hypergeometric function, respectively.
Nevertheless, in what follows, we will continue to discuss the computation of observables considering $p_T$ as the random variable.

\subsection{Average of transverse momentum}
The first moment of $p_T$ of the three different approaches are given by
\begin{subequations}
\label{eq:avgpT}  
\begin{eqnarray}
        \avg{\mathcal{P}_T}_\text{th} &=& T_\text{th} ,
        \label{eq:avgpTth}
    \\
        \avg{\mathcal{P}_T}_\text{Hag} 
        &=& \frac{m}{m-2} T_\text{Hag},        
    \\
         \avg{\mathcal{P}_T}_U 
     &=& \frac{4 - 2q}{5 - 3q} T_U.
    \end{eqnarray} 
\end{subequations}
The transverse momentum averages are
\begin{subequations}
\label{eq:meanpT}
\begin{eqnarray}
   \avg{p_T}_\text{th} &=& 2T_\text{th} ,
        \label{eq:meanpTth}
            \\
        \avg{p_T}_\text{Hag}  
        &=& \frac{2m}{m-3} T_\text{Hag},
         \label{eq:32b}
    \\
        \avg{p_T}_U 
     &=& \frac{(q-1)(3q-5)}{(2-q)(2q-3)} \left( \frac{\Gamma\left( \frac{1}{q-1}\right) }{\Gamma\left( \frac{1}{q-1} - \frac{1}{2}\right)} \right)^2 T_U .\qquad
    \label{eq:32c}
    \end{eqnarray}
\end{subequations}

\begin{figure}[ht]
    \centering
    \includegraphics[scale=0.5]{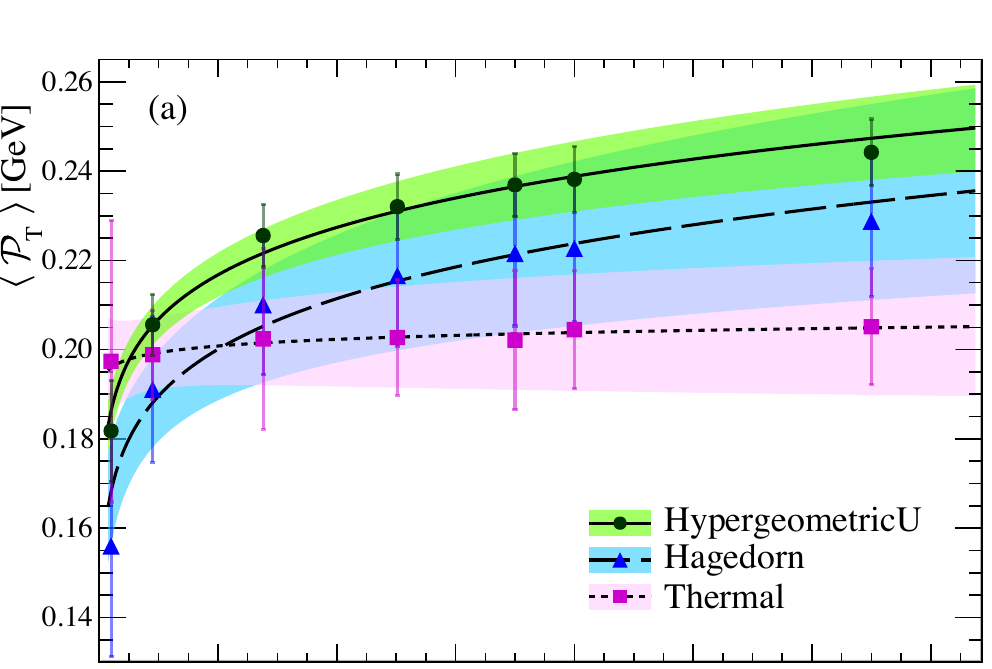}
    \includegraphics[scale=0.5]{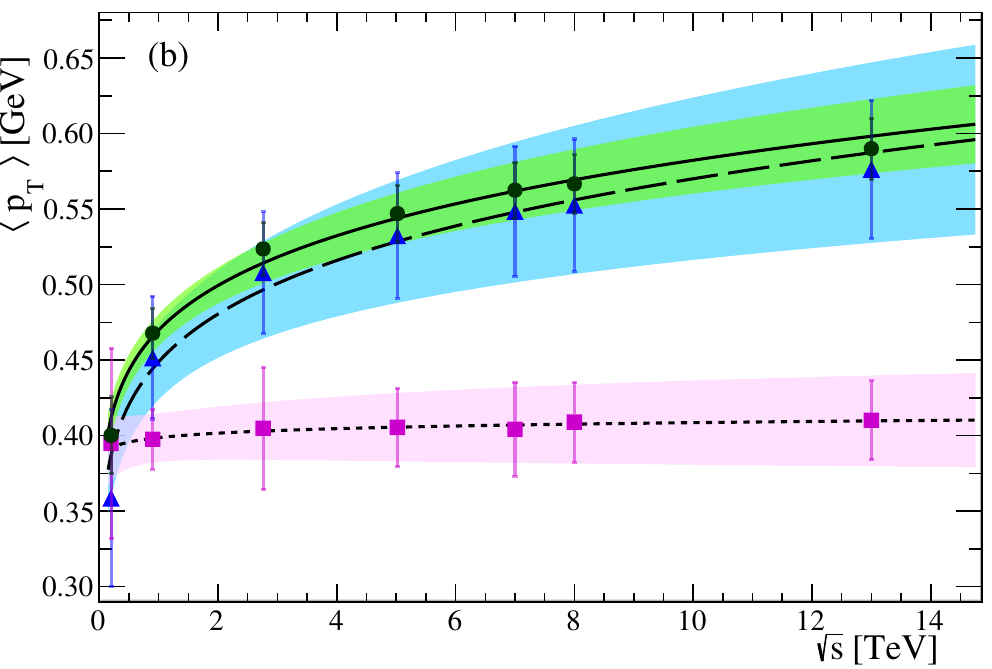}    
    \caption{(a) $\avg{\mathcal{P}_T}$ and (b) $\avg{p_T}$ of the transverse momentum distribution as a function of the center of mass energy for pp collisions for the three different fitting functions with their corresponding parameter dependence. Lines, figures, and colors are the same as in Fig.~\ref{fig:Temp}.}
    \label{fig:pt1}
\end{figure}

We recall that the fit parameters $q$, $\sigma$, $m$, $p_0$, and $T_\text{th}$ exhibit a particular dependence with the center of mass energy for the case of minimum bias pp collisions. This behavior can be incorporated into the average of $p_T$ by plugging Eqs.~\eqref{eq:allfit},~\eqref{eq:Hagpar}, and~\eqref{eq:Tfit} into the Eqs.~\eqref{eq:avgpT}, and~\eqref{eq:ptnratio} for $n=1$. Figure \ref{fig:pt1} shows the behavior of $\avg{\mathcal{P}_T}$ and $\avg{p_T}$ as a function of the center of mass energy of minimum bias pp collisions for the three different approaches with their corresponding estimation discussed above.

It is worth mentioning that the average of $p_T$ is proportional to the thermal temperature in the three approaches, given by simple combinations of $q$ and $m$ for the $U$ and Hagedorn functions, respectively. These parameters are the exponents that modulate the hard part of the TMD. Equations~\eqref{eq:32b} and~\eqref{eq:32c} lead to an enhancement of the $\avg{p_T}$ when compared with the thermal function, as seen in Fig.~\ref{fig:pt1}, but they recover Eq.~\eqref{eq:meanpTth} in the limit $q\to 1$ and $m \to \infty$.

\begin{figure}[ht]
    \centering
    \includegraphics[scale=0.25]{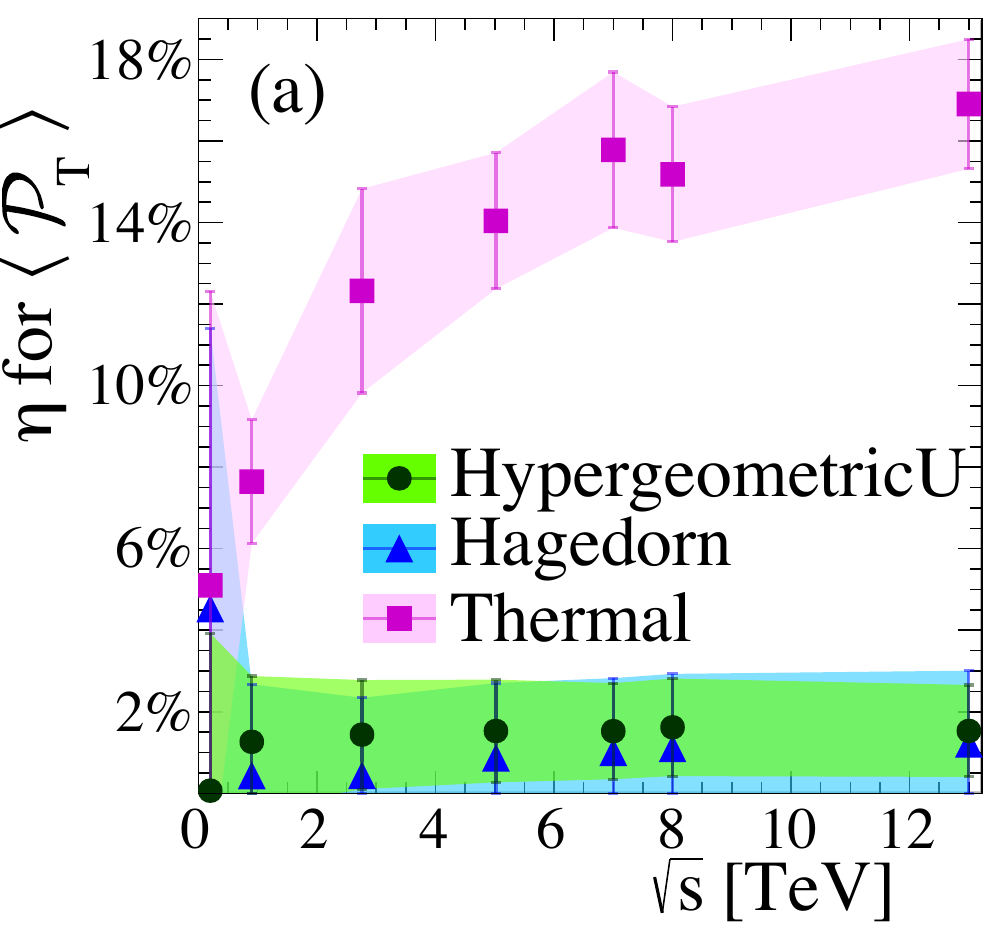}
    \includegraphics[scale=0.25]{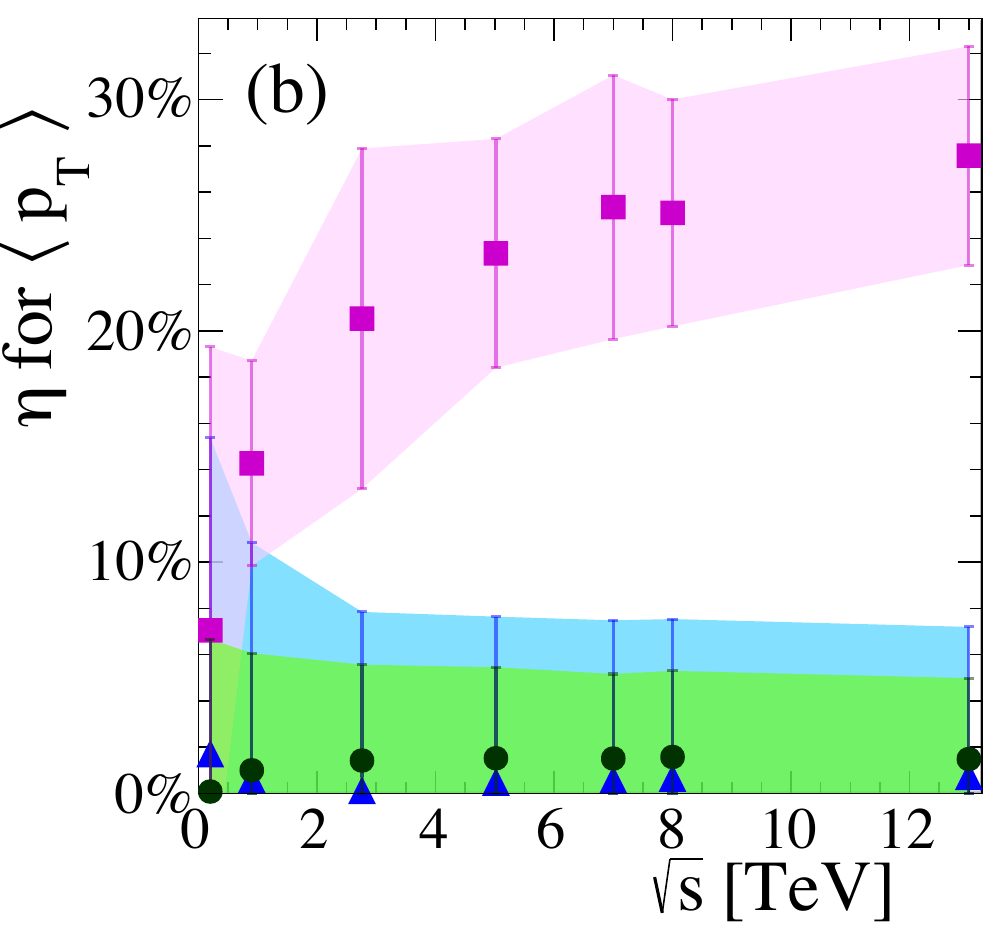}
    \caption{Absolute percentile deviation of (a) $\avg{\mathcal{P}_T}$ and (b) $\avg{p_T}$ with the one computed directly from the experimental data for the confluent hypergeometric $U$ (circles), Hagedorn (triangles) and Thermal (squares) fitting functions. Figures and colors are the same as in Fig.~\ref{fig:Temp}.}
    \label{fig:comparisonpT}
\end{figure}

Additionally, we can compare the average of the transverse momentum statistics computed directly from the experimental TMD data. Thus, the $n$-moment is calculated as discussed above, but now we compute the $I_n$ integrals as follows:
\begin{equation}
    I_n^\text{hist} = \sum_k p_{Tk}^n \text{TMD}_k \Delta p_{Tk},
\end{equation}
where $k$ is the bin number, $p_{Tk}^n$ is the conservative $p_T$ value of the $k$-bin, $\text{TMD}_k$ is the TMD value reported for the $k$-bin, and $\Delta p_{Tk}$ is the bin width.
We also added the superscript hist to differentiate the $I_n$ integrals computed from the TMD histogram.
Moreover, to compare the predictions of the fitting function, we define the absolute percentage of deviation as
\begin{equation}
    \eta=\frac{|\avg{\mathcal{P}_T}^\text{hist}-\avg{\mathcal{P}_T}^\text{trunc}|}{\avg{\mathcal{P}_T}^\text{hist}},\label{eq:deviation}
\end{equation}
where $\avg{\mathcal{P}_T}^\text{hist}=I_1^\text{hist}/I_0^\text{hist}$, and
\begin{equation}
   \avg{\mathcal{P}_T}^\text{trunc} = \frac{1}{I_0^\text{hist}} \int_\mathcal{R} p_T \text{TMD} dp_T, 
\end{equation}
with $\mathcal{R}$ being the $p_T$ range reported by experiments.
Similarly, the absolute percentage of deviation of the $p_T$ average by replacing $\avg{\mathcal{P}_T}$ with $\avg{p_T}$ in Eq.~\eqref{eq:deviation}.
Figure~\ref{fig:comparisonpT} shows our comparison for the first moment and the average of $p_T$. Notice the agreement between the estimations of the Hagedorn and $U$ fitting function and the value computed from the experimental data.

\subsection{Variance of the transverse momentum}
\begin{figure}[ht]
    \centering
    \includegraphics[scale=0.5]{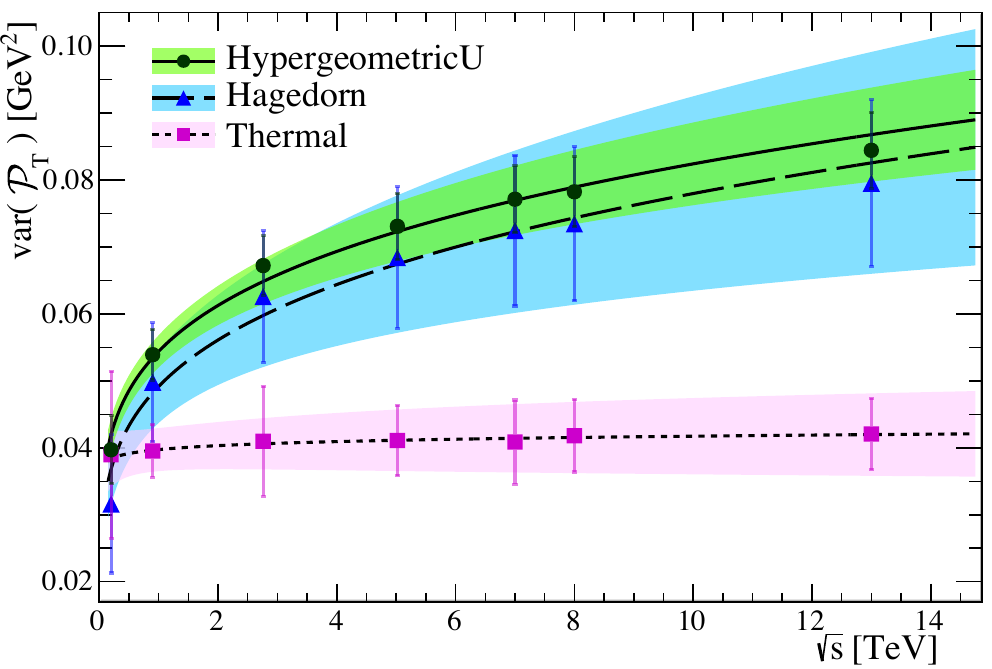}
    \caption{Variance of the TMD as a function of the center of mass energy for minimum bias pp collisions for the three different approaches. Lines, figures, and colors are the same as in Fig.~\ref{fig:Temp}.}
    \label{fig:varpT}
\end{figure}

The variance of the TMD is immediately calculated as var$(\mathcal{P}_T)=\langle \mathcal{P}_T^2 \rangle-\langle \mathcal{P}_T \rangle^2$ for the three approaches considered
\begin{subequations}
\label{eq:varpT}
\begin{eqnarray}
        \text{var}_\text{th}(\mathcal{P}_T) &=& T_\text{th}^2 ,
        \label{eq:varpTth}
   \\
        \text{var}_\text{Hag}(\mathcal{P}_T) 
        &=& \frac{m^2(m+1)}{(m-3)(m-2)^2} T_\text{Hag}^2,    
        \\
         \text{var}_U(\mathcal{P}_T) 
     &=&  F(q) T_U^2,
    \end{eqnarray}
with $F(q) = \frac{2(q-1)}{3-2q} \left( \frac{  \Gamma\left( \frac{1}{q-1} \right) }{ \Gamma\left( \frac{1}{q-1} - \frac{1}{2} \right) }\right)^2 - \left(\frac{4 - 2q}{5 - 3q} \right)^2 $.   

\end{subequations}
The dependence of $q$, $\sigma$, $m$, $p_0$, and $T_\text{th}$ with the center of mass energy are considered into the Eqs.~\eqref{eq:varpT} via Eqs.~\eqref{eq:allfit},~\eqref{eq:Hagpar}, and~\eqref{eq:Tfit}. Figure \ref{fig:varpT} shows the variance of the TMD as a function of the center of mass energy of minimum bias pp collisions for the three different approaches with their corresponding estimation. Similarly to the $\avg{\mathcal{P}_T}$ case, the variance is proportional to the squared thermal temperature in the three cases, given by combinations of the exponent parameters of Hagedorn and $U$ fitting functions. The expressions~\eqref{eq:varpT} reveal that the width of the TMD for the Hagedorn and the $U$ are larger than the thermal's, as seen in Fig.~\ref{fig:varpT}. 

It is worth mentioning that the computation of $\avg{p_T^2}$ is crucial for the phenomenology calibration of some models, like the Color String Percolation Model \cite{Braun:2015eoa,bautista2019string}. In this model, the average of $p_T^2$ and the multiplicity of the produced charged particles comes from the color interaction between strings. It has been shown that overlapping the color string leads to a suppression of the color field.
An immediate consequence is that the clusters of strings produce fewer particles per string but enhance their transverse momentum. Finally, the comparison between the estimated $\avg{p_T^2}$ makes the CSPM can be compared with the experimental data \cite{jhony-area, ramirez2021interacting, garcia2022percolation,alvarado2023structure}.

\subsection{Kurtosis of the TMD}
\begin{figure}[ht]
    \centering
    \includegraphics[scale=0.5]{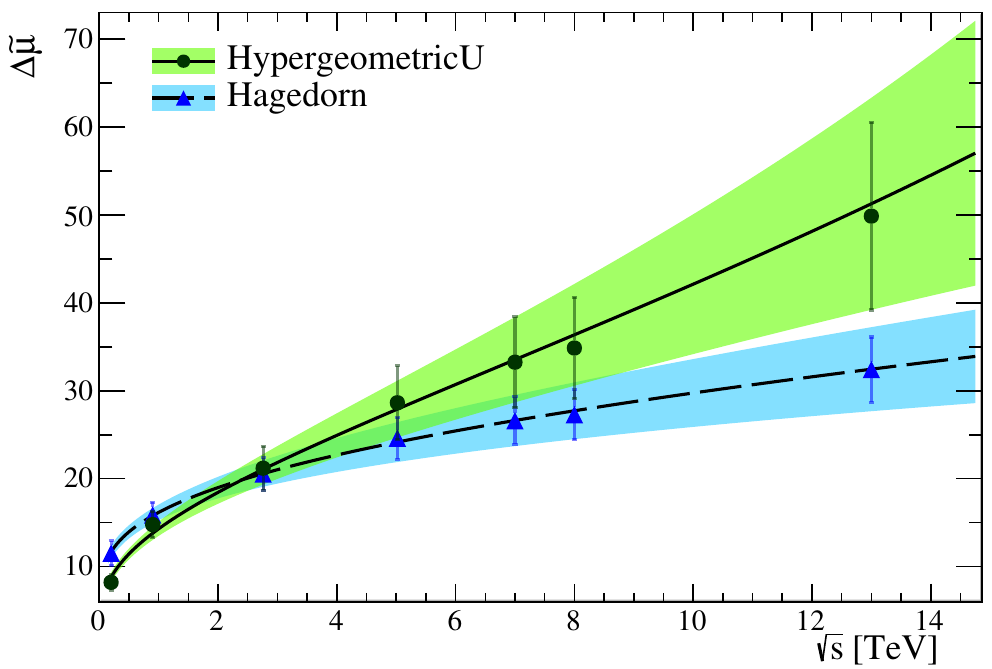}
    \caption{Excess of kurtosis calculated with respect to the thermal as a function of the center of mass energy for pp collisions. Lines, figures, and colors for the Hagedorn and $U$ fitting functions are the same as in Fig.~\ref{fig:Temp}.}
    \label{fig:kurtpT}
\end{figure}

The kurtosis is calculated as usual:
\begin{equation}
      \widetilde{\mu} =  
    \frac{ \avg{\mathcal{P}_T}^4 }{ \left[ \text{var}(\mathcal{P}_T) \right]^2 } 
    \left[ \frac{\avg{\mathcal{P}_T^4}}{\avg{\mathcal{P}_T}^4} - 4  \frac{\avg{\mathcal{P}_T^3}}{\avg{\mathcal{P}_T}^3} + 6 \frac{\avg{\mathcal{P}_T^2}}{\avg{\mathcal{P}_T}^2} - 3\right],  
    \label{eq:kurt}
\end{equation}
which for the thermal case is exactly $9$. For the Hagedorn and $U$ functions, we substitute the needed moments from Eqs.~\eqref{eq:nptHag}, and~\eqref{eq:nptU} into the Eq.~\eqref{eq:kurt}. We also consider the dependence of the fitting parameters on the center of mass energy, as discussed in Sec.~\ref{sec:TMDfit}. Figure~\ref{fig:kurtpT} shows the excess of kurtosis, defined as $\Delta \widetilde{\mu} =  \widetilde{\mu} - 9$, for the Hagedorn and hypergeometric $U$ fitting functions.

Notice that both the Hagedorn and $U$ fitting functions reveal that their descriptions contain more information about heavy tails since $\Delta \widetilde{\mu} >0$.  
Furthermore, $\Delta \widetilde{\mu}$ increases as the center of mass energy rises. In fact, the $q$-Gaussian fluctuations induce a TMD with more information in the tail than the Hagedorn, despite that the latter is a QCD based function. 
Remarkable, the $U$ distribution encodes information related to both soft and hard scales.

\section{Shannon entropy and heat capacity}\label{sec:Shannon}

Let us delve into a fundamental concept in information theory, the Shannon entropy. It provides a way for quantifying the uncertainty and information of the TMD \cite{shannon1948mathematical}. This observable can shed light on the characteristics of the final state particles of collision systems. 
Since the temperature-like parameter $T$ is extracted from the TMD in each approach, the natural way of computing the Shannon entropy is by considering the normalized TMD as the probability density function of the random variable $p_T$, as usually done in the generalized ensemble theory \cite{niven2010jaynes,langen2015experimental}. Then,
the Shannon entropy is computed as \cite{shannon1948mathematical}
\begin{equation}
    \mathcal{H} =-\int_0^\infty 
    (\text{TMD}/I_0 ) \ln \left[ \text{TMD} /I_0   \right] dp_T,
  \label{eq:shannonH}  
\end{equation}
where $I_0$ is the normalization constant given by Eq.~\eqref{eq:In}.

\begin{figure*}[ht]
\begin{center}  
    \includegraphics[scale=1]{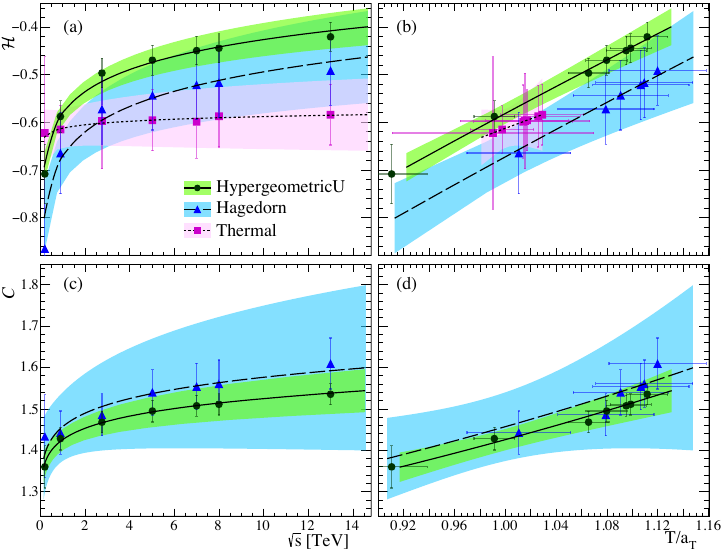} 
\end{center}
    \vspace{-10pt}
\caption{ Shannon entropy dependence on the (a) center of mass energy and (b) temperature. Heat capacity as a function of the (c) center of mass energy and (d) temperature scaled by $a_T$. The shaded regions correspond to the uncertainty propagation of the corresponding equations. Lines, figures, and colors for the thermal, Hagedorn, and $U$ fitting functions are the same as in Fig.~\ref{fig:Temp}.
}
\label{fig:Shannon}
\end{figure*}
    \vspace{-30pt}

The Shannon entropy~\eqref{eq:shannonH} can be expressed in terms of elementary functions for the thermal and Hagedorn fitting functions. For these cases, we obtain
\begin{subequations}
\label{eq:Shannon}
\begin{eqnarray}
    \mathcal{H}_\text{th} &=& 1+\ln(T_\text{th}).
\label{eq:ShannThermal}
\\
    \mathcal{H}_\text{Hag} 
    &=& \frac{m}{m-1} + \ln \left( \frac{m}{m-1} \right) + \ln (T_\text{Hag}),
\label{eq:ShannHag}
\end{eqnarray}
\end{subequations} 
respectively.

On the other hand, the Shannon entropy of the confluent hypergeometric function is explicitly given by
\begin{equation}
    \mathcal{H}_U =-\int_0^\infty 
    ( U(a, b, z)/I_0 ) \ln \left[ U(a, b, z) /I_0   \right] dp_T,
  \label{eq:ShannU}  
\end{equation}
with
\begin{equation}
    I_0 =
    \frac{\sigma}{(2-q) \Gamma(a)} \sqrt{\frac{ q-1}{2} }  ,
\end{equation}
which can be rewritten as
\begin{equation}
\mathcal{H}=\ln(I_0) + \frac{\mathcal{H}_1}{I_0},
\end{equation}
where 
\begin{equation}
    \mathcal{H}_1 = -\int_0^\infty 
     U(a, b, z)   \ln \left[ U(a, b, z) \right] dp_T. \label{eq:H1}
\end{equation}
As far as we know, the Eq.~\eqref{eq:H1} cannot be solved analytically. Then, we computed $\mathcal{H}_1$ by using numerical methods. Figure \ref{fig:Shannon} shows our estimations of the Shannon entropy for minimum bias pp collisions as a function of the center of mass energy and the corresponding temperature for each approach.

We also compute the heat capacity using its thermodynamic definition \cite{Mandal_2013} 
\begin{equation}
    C = T\frac{d\mathcal{H}}{dT} .
    \label{eq:heatdef}
\end{equation}
In this context, Eq.~\eqref{eq:heatdef} is a measure of how much “heat” is necessary to “warm” the TMD. “Heating” the TMD must be understood as a global change of the TMD shape, flattening the soft part together with an enhancement of the TMD tail.

To compute Eq.~\eqref{eq:heatdef}, we must take into account that the fitting parameters may depend on the temperature. In these cases, the computation of the heat capacity must be done using the chain rule. 
In particular, for the thermal case, we found $C_\text{th} = 1$. 
In the case of the Hagedorn fitting function, the heat capacity is given by
\begin{equation}
C_\text{Hag} = 1 + T_\text{Hag} \frac{1-2m}{m(m-1)^2} \frac{d m}{d T_\text{Hag}}, \label{eq:CHag}  
\end{equation}
where the derivative $dm/dT_\text{Hag}$ for minimum bias pp collisions is computed through Eq.~\eqref{eq:mfit} and using the inverse relation of the temperature with the center of mass energy in Eq.~\eqref{eq:Tfit}, which reads
\begin{equation}
\frac{dm}{dT_\text{Hag}}=\frac{c_m m(T_\text{Hag})}{c_T T_\text{Hag}}.
\label{eq:dmdT}
\end{equation}
Plugging Eq.~\eqref{eq:dmdT} into \eqref{eq:CHag}, the heat capacity is
\begin{equation}
C_\text{Hag} = 1 + \frac{c_m(1-2m(T_\text{Hag}))}{c_T(m(T_\text{Hag})-1)^2},
\end{equation}
with $m(T_\text{Hag})=a_m (T_\text{Hag}/a_{T})^{c_m/c_{T}}$.

On the other hand, for the calculation of the heat capacity of the confluent hypergeometric $U$ function, we start replacing $\sigma$ in favor of $T_U$ through Eq.~\eqref{eq:Tthermal}.
Thus, the normalization constant $I_0$ and the $z$ parameter (see~\eqref{eq:abz}) in the third argument of the $U$ functions are rewritten as follows 
\begin{eqnarray}
  I_0  &=&  \sqrt{\pi}\frac{(q-1)\Gamma\left( a+\frac{1}{2} \right)}{(2-q)\Gamma(a)}T_U=I_{0q}(q)T_U,  
\end{eqnarray}
\begin{eqnarray}
z  &=&  \left( \frac{\Gamma(a)}{\Gamma\left( a+\frac{1}{2} \right)}  \right)^2 \frac{p_T^2}{2T_U^2}.
\end{eqnarray}
Therefore, the heat capacity for the $U$ fitting function is
\begin{equation}
\begin{split}
    C_U= &
\left( 1-\frac{\mathcal{H}_1}{I_0}  \right) \left( 1+T_U\frac{I_{0q}'}{I_{0q}} \frac{dq}{dT_U} \right) \\ &
- 
\frac{T_U} {I_0} \int_0^\infty (\ln U +1) \frac{dU}{dT_U} dp_T. \label{eq:CU}
\end{split}
\end{equation}
In Eq.~\eqref{eq:CU}, the remaining integral is 
\begin{widetext}
\begin{equation}
\begin{split}
\int_0^\infty (\ln U+1)  \frac{dU}{dT_U} dp_T
=&  \frac{\sqrt{2\pi}a(q-2) \Gamma\left( a+\frac{1}{2}  \right)}{\left( a-\frac{1}{2} \right)^2(q-3) \Gamma(a)} \left\{ \left( \frac{\Gamma\left( a+\frac{1}{2}  \right)}{\Gamma(a)}  \right)^2- \left( a +\frac{1}{2}\right)^2 \left[ \psi^{(0)}\left( a +\frac{1}{2}\right)-\psi^{(0)}(a)  \right] T_U \frac{dq}{dT_U} \right\} 
\\& - \left( a +\frac{1}{2}\right) \frac{dq}{dT_U}\int_0^{\infty} \left[ \ln U(a, 1/2, z) +1 \right] U^{(1, 0, 0)}\left(a, 1/2, z\right) dp_T+\frac{a}{T_U^2}\left(   \frac{\Gamma(a)}{\Gamma(a+\frac{1}{2})}  \right)^2
\\&  \times\left( \frac{1}{T_U}-\left[\psi^{(0)}\left( a +\frac{1}{2}\right)-\psi^{(0)}(a)\right] \frac{dq}{dT_U} \right) \int_0^\infty p_T^2 U(a+1, 3/2, z)\ln U(a, 1/2, z) dp_T,
\end{split}
\end{equation}
\end{widetext}
where $\psi^{(0)}(x)$ is the zeroth order polygamma function. We also added the superscript to the $U$ function to denote their first derivative with respect to the first argument. Notice that there are two remaining integrals, which are done by means of numerical methods.
Similarly to the case of the Hagedorn fitting function, $q$ is expressed as a power law as a function of $T_U$, i.e., $q(T_U)=a_q(T_U/a_T)^{c_q/c_T}$, and its derivative with respect to the temperature is
\begin{equation}
\frac{dq}{dT_U}=\frac{c_q}{c_T} \frac{q(T_U)}{T_U}.
\end{equation}

The heat capacity of the three different schemes as a function of the center of mass energy and the temperature is plotted in Fig.~\ref{fig:Shannon}.

\section{Conclusions}\label{sec:conclusions}

In this work, we discussed the statistics of three different fitting functions that describe the TMD data, namely, the thermal $p_T$-exponential, the confluent hypergeometric, and the Hagedorn. The former arises from the string tension fluctuation in the QCD color string picture, meanwhile, the latter is a power law inspired by the foundations of QCD.
All of them predict a temperature parameter at the low $p_T$ regime.

The temperature estimated for minimum bias pp collisions as a function of the center of mass energies reflects the physical motivation of the different approaches. The thermal distribution adequately describes the soft part because it assumes an exponential decay (similar to a Boltzmann distribution), resulting in overestimating the temperature for the complete TMD. 
On the contrary, the power law proposed by Hagedorn establishes a description of the hard processes, leading to the heavy tail of the spectrum. This means that the thermal temperature may not precisely incorporate the soft part of the TMD. In fact, Hagedorn suggests that their fit must be performed in the $p_T$ interval from $0.3$ to $10$ GeV \cite{Hagedorn:1983wk}.
On the other hand, we must emphasize that the confluent hypergeometric $U$ adequately combines the information of the soft and hard scales to predict the temperature.

We also discuss the statistics of the normalized TMD by computing the moments of the distribution and, thus, the variance and kurtosis for the three different approaches. This analysis lets us distinguish the particularities of each fitting function. For example, the Hagedorn and $U$ functions reveal more dispersion than the thermal one because of the information coming from the heavy tail. In all cases, we found an increasing trend on the first moment and variance with the center of mass energy as seen in Figs.~\ref{fig:pt1} and~\ref{fig:varpT}. Moreover, the heavy tail absence in the thermal distribution leads to a constant kurtosis, which was taken as a reference to measure the excess of kurtosis in the Hagedorn and $U$ distributions, both increase as the center of mass energy rises, highlighting that the $U$ grows more substantially (see Fig.~\ref{fig:kurtpT}).  This means the TMD derived from the $q$-Gaussian string tension fluctuations contains more information in the tail than the Hagedorn approach. This is important because the $U$ fitting function adequately reproduces the power law behavior associated with the QCD hard processes from the color string picture.

In addition, we compare the average of the transverse momentum estimated from the fitting function and the computed from the experimental data. It was found that the Hagedorn and the $U$ functions precisely reproduce the value of $\avg{p_T}^\text{hist}$, but the predictions of the thermal distribution considerably deviate (15\%-25\%).

Other observables that we computed are the Shannon entropy and the heat capacity.
The Shannon entropy increases as the center of mass energy grows. This is consistent with the TMD variance, which exhibits a similar behavior. This happens because the probability of observing particles with high $p_T$ rises with an increment on the center of mass energy. Then, the TMD suffers a global widening and an enhancement of its tail. 
Moreover, we observed quite differences between the entropy computed for the Hagedorn and hypergeometric confluent $U$ functions. These subtle deviations may come from the shape of the TMD at very low $p_T$, which can be inferred from the temperature estimated by each model. 

Moreover, the computation of the heat capacity for the thermal fitting function reveals that the system does not change its requirements to \emph{heat up}. 
From a classical thermodynamics point of view, this means that collision systems described by a thermal distribution resemble an ideal gas of monoatomic (or rigid diatomic) molecules. 
On the contrary, for the Hagedorn and $U$ functions, the heat capacity grows as the temperature does, similar to a thermodynamic system that can manifest other degrees of freedom when heating.
This implies that, to \emph{heat up} the collision system, it is necessary to reach an increasingly higher center of mass energies. This is a direct consequence of the heavy tailed TMD, since it requires not only \emph{heating} the thermal part but also the hard one for the discussed minimum bias pp collisions. From our results, we infer that the higher the TMD temperature, the more energetic collision is required. This observation is consistent with the analysis of the experimental data of the temperature saturation as a function of the center of mass energy (see Fig.~\ref{fig:Temp}).

This work can be extended in several ways. For instance, it would be interesting to analyze the TMD and compute the Shannon entropy and heat capacity of pp collisions as a function of the multiplicity, heavy ion collisions, production of identified particles, and other processes.
Part of these results are currently under discussion, and we will report our findings in a future paper.

\section*{Acknowledgments}
This work has been funded by the projects PID2020-119632GB-100 of the Spanish Research Agency, Centro Singular de Galicia 2019-2022 of Xunta de Galicia and the ERDF of the European Union.
This work was funded by Consejo Nacional de Humanidades, Ciencias y Tecnologías (CONAHCYT-México) under the project CF-2019/2042,
graduated fellowship grant number 1140160, and postdoctoral fellowship grant numbers 289198 and  645654.
J. R. A. G. acknowledges financial support from Vicerrectoría de Investigación y Estudios de Posgrado (VIEP-BUAP).

\bibliography{refprd.bib}

\end{document}